\documentclass[12pt]{article}
\usepackage[utf8]{inputenc}    
\usepackage{amsmath}
\usepackage{amsthm}
\usepackage{amssymb}
\usepackage{graphicx}
\usepackage{epstopdf}
\usepackage{inputenc}
\usepackage{geometry}
\usepackage[hyphens]{url}
\usepackage{hyperref}

\usepackage{algorithmicx}
\usepackage{algorithm}
\usepackage{mdframed}

\usepackage{float}
\usepackage[noend]{algpseudocode}
\algnewcommand\algorithmicinput{\textbf{Input:}}
\algnewcommand\INPUT{\item[\algorithmicinput]}
\algnewcommand\algorithmicoutput{\textbf{Output:}}
\algnewcommand\OUTPUT{\item[\algorithmicoutput]}

\usepackage{authblk}
\usepackage{makeidx} 
\usepackage{tabularx,ragged2e,booktabs,caption}

\usepackage{authblk}
\usepackage[nottoc]{tocbibind}
\usepackage{makecell}
\usepackage{color}
\usepackage[disable]{todonotes} 
\usepackage[
    backend=biber,
    style=bwl-FU,
  ]{biblatex}
 
\usepackage{pgfplots}
\pgfplotsset{width=10cm,compat=1.9}
\definecolor{darkblue}{rgb}{0.0,0.0,0.3}
\numberwithin{equation}{section}

\hypersetup{
    colorlinks=true,
    allcolors=black,
    linktocpage=true,
    bookmarksopen=true,
  }

\newcommand{\commentOut}[1]{}

\makeatletter
\def\BState{\State\hskip-\ALG@thistlm}
\makeatother

\title{Auditing Indian Elections}
\author[1]{Vishal Mohanty}
\author[2]{Nicholas Akinyokun}
\author[3]{Andrew Conway}
\author[2]{Chris Culnane}
\author[4]{Philip B.\ Stark}
\author[2]{Vanessa Teague}
\affil[1]{Department of Computer Science and Engineering\\Indian Institute of Technology, Madras\\E-mail: \texttt{cs15b039@smail.iitm.ac.in}\\Tel. \texttt{+91 9840298064}\\
This work was done while visiting the University of Melbourne}
\affil[2]{ School of Computing and Information Systems\\
University of Melbourne\\
\texttt{\{christopher.culnane, oakinyokun, pstuckey, vjteague\}@unimelb.edu.au}
}
\affil[3]{aindaud@greatcactus.org}
\affil[4]{Department of Statistics\\
 University of California, Berkeley\\
 \texttt{stark@stat.berkeley.edu}
 }

\begin{document}
\date{\today}
\maketitle

\begin{abstract}
Indian Electronic Voting Machines (EVMs) will be fitted with printers that produce Voter-Verifiable Paper Audit Trails (VVPATs) in time for the 2019 general election.  VVPATs provide evidence that each vote was recorded as the voter intended, without having to trust the perfection or security of the EVMs.
However, confidence in election results requires more: VVPATs must be preserved inviolate and then actually used to check the reported election result in a trustworthy way that the public can verify.  A full manual tally from the VVPATs could be prohibitively expensive and time-consuming; moreover, it is difficult for the public to determine whether a full hand count was conducted accurately. 
We show how \emph{Risk-Limiting Audits} (RLAs) could provide high confidence in Indian election results.
Compared to full hand recounts, RLAs typically require manually inspecting far fewer VVPATs when the outcome is correct, and are much easier for the electorate to observe in adequate detail to determine whether the result is trustworthy.

We discuss how two RLA strategies, \emph{ballot-level comparison} and \emph{ballot polling}, can be applied to General Elections in India. 
Our main result is a novel method for combining RLAs in constituencies to obtain an  RLA of the overall parliamentary election result.  

\end{abstract}



\section{Introduction}\label{secIntroduction}

Since Electronic Voting Machines (EVMs) were introduced in India in the 1999 elections, there have been questions about their transparency and trustworthiness; a number of security vulnerabilities have been documented (\cite{wolchok2010security}). In 2013, the Indian Supreme Court ruled that all EVMs in Indian General Elections must be equipped with printers providing Voter-Verifiable Paper Audit Trails (VVPATs, \cite{IndiaSCVVPATs}). 
The Election Commission of India has introduced VVPAT-equipped EVMs in several constituencies and has promised that all EVMs used in the upcoming 2019 General Election will have VVPAT printers. 

VVPATs allow each voter to verify that his or her 
intended selections are correctly printed on a paper record, which is collected in a separate container called the \emph{VVPAT box}. 
VVPATs provide a way to check and correct election results, for instance, if there is a legal demand by a candidate, or for routine checks of election tabulation accuracy---audits.
VVPATs could be manually recounted to check the electronic results, but that is labor-intensive and time-consuming.  
We show how auditing a random sample of VVPAT records can justify confidence in election results without a full manual tally.  
Auditing a VVPAT means manually inspecting the paper record to see the voter preferences it shows. Different auditing strategies---for instance, ballot-level comparison audits and ballot-polling audits---use that information differently, as described below. 

The Election Commission (EC) of India is taking steps to increase the transparency of the Indian electoral system. 
In a recent report,\footnote{\url{http://indianexpress.com/article/india/ec-to-tally-paper-trail-slips-with-evms-in-5-pc-booths-in-each-assembly-seat-4737936/}} the EC decided to tally the paper trail slips and compare them with the electronic result provided by the EVMs in $5\%$ of the booths in each Assembly seat district, selected randomly. 
This effort, while well-intentioned, does not suffice to give strong evidence that election results are correct. 
In this paper, we show rigorous ways of attaining well-defined confidence levels. 

Suitable post-election audits may justify confidence of voters, candidates, and parties that election results are correct. One type of post-election audit is a Risk-Limiting Audit (RLA), which either develops strong statistical evidence that the reported outcome is correct, or corrects the results (by conducting a full manual tally of a reliable paper trail). 
Here, ``outcome'' means the set of reported winners of the contests, not the exact vote tallies.
To ensure that the tallies are correct to the last vote is prohibitively expensive, if not impossible; conversely, to ensure that the reported winners really won seems like the lowest reasonable standard for accuracy.

Before a RLA commences, the \emph{risk limit} (denoted $\alpha$) must be chosen; ideally, it is set in legislation or regulation, so that auditors cannot manipulate the level of scrutiny a contest gets by adjusting the risk limit.
The risk limit is the maximum probability that the audit will fail to correct the reported election outcome, on the assumption that the reported outcome is wrong.
The risk limit is a worst-case probability that makes no assumption about \emph{why} the outcome is wrong, \emph{e.g.}, it could be because of accidental error, procedural lapse, bugs, misconfiguration, or malicious hacking by a strategic adversary who knows how the audit will be conducted. 
RLAs assume that the paper ballots reflect the correct outcome, \emph{i.e.} that a full manual tally of the paper trail would show who really won.
A RLA of an unreliable paper trail is ``security theater.'' Hence, there need to be procedures (called \emph{compliance audits} by \cite{benaloh2011soba,LindemanStark2012,stark2012evidence,stark2018}) to ensure that the paper trail is complete and intact before the RLA begins. 

This paper shows how two types of RLAs can be used with Indian elections: transitive ballot-level comparison audits and ballot-polling audits.
Ballot-level comparison audits require a ``commitment'' to the interpretation of each ballot in a way that allows that interpretation to be compared to a human reading of voter intent directly from the paper ballot.  
Ballot-polling audits do not require knowing how individual ballots were interpreted.  
Ballot-level comparison audits are more efficient in the sense that they generally involve inspecting fewer ballots to attain the same risk limit when the reported outcome is correct.  
However, they require more setup. 
As discussed in Section~\ref{secApplying}, they may require a voting system that can export its interpretation of individual ballots in a way that can be matched to the corresponding paper, or may require sorting the physical ballots or VVPATs before the audit, according to the votes they (reportedly) show. 

Our main contribution is to develop RLAs for a new social choice function---Indian parliamentary majorities---with procedures suited to the logistics of Indian elections.
To verify the overall election outcome we need to verify that the party/coalition reported to have been elected to form the government actually won.
That generally requires less auditing than confirming the winner in every constituency. 
The method we develop splits the responsibility of the auditing among various constituencies in a way that the combined result gives higher confidence in the correctness of the overall parliamentary outcome than each constituency would have in its results alone. This procedure is discussed in Section~\ref{secOverall}. 
Our methods apply to any parliamentary democracy, but the computations
are particularly simple when all constituencies have equal weight.

\section{Background}
\subsection{Indian Elections}
Indian General elections are held Quinquennially to elect the Lok Sabha (Lower House of the Parliament). The country is divided into 543~constituencies, each represented by one person elected to the Lok Sabha. 
Elections at the constituency level are  \emph{plurality} contests: the person who gets the most votes wins. 
Candidates at the constituency level typically belong to some political party, but can be unaffiliated with any party. 
At the parliamentary level, the party that gets the \emph{majority} of the seats forms the government. 
If no party has a majority, parties may form coalitions to attain a majority.
Coalitions can be formed before or after elections, although before is more common.  Elections are conducted in phases spread over a month. 
Each phase consists of single-day elections in a subset of constituencies, typically grouped by geography.

\subsection{Related work on election auditing}\label{secRelated}

RLAs are procedures that guarantee a minimum chance of conducting a full manual tally of the voter-verifiable records when the result of that tally would belie the reported outcome.
They amount to a statistical test of the \emph{null hypothesis} that the election outcome is wrong, at significance level $\alpha$, the \emph{risk limit}, chosen in advance.  
A RLA continues to examine more ballots until the null hypothesis is rejected at significance level $\alpha$, allowing one to conclude with known confidence that the outcome is in fact correct---or until there has been a complete manual tally to set the record straight. 
The risk limit is the largest chance that the audit will \emph{not} require a full manual tally of the paper records if the electoral outcome according that tally would differ from the reported electoral outcome.

RLAs were introduced by
 \cite{Stark2008a}, but were not so named until \cite{Stark2009b}.
The first RLAs were conducted in California in 2008 (\cite{HallEtal2009}).
RLAs have been conducted in California, Colorado, Indiana, Michigan, New Jersey, Ohio, Virginia, and Denmark.
RLAs have been developed for a variety of social choice functions and for  variety of sampling strategies (unstratified sampling of individual ballots or batches, with or without replacement; stratified sampling with and without replacement, Bernoulli sampling) and auditing strategies (\emph{batch-level comparisons}, \emph{ballot-level comparisons}, and \emph{ballot polling}).

\emph{Ballot-polling audits} 
(\cite{Lindeman2012,LindemanStark2012,bernhardEtal2019})
do not require knowing how the system interpreted individual ballots nor how it tallied the votes on subsets of ballots.
They directly check whether the reported winner(s) received more votes than the reported loser(s) by sampling and manually interpreting individual ballots.
To draw a random sample of ballots typically involves a \emph{ballot manifest}, which describes how the physical ballots are organized: the number of bundles, the labels of the bundles, and the number of ballots in each bundle.
(However, see \cite{bernhardEtal2019}.)

The BRAVO ballot-polling method (\cite{Lindeman2012}) uses Wald's sequential probability ratio test (\cite{Wald1945}) to test a collection of null hypotheses, namely, that any loser in fact tied or beat any winner.
For each (winner, loser) pair, the audit tests the hypothesis that the loser got as many or more votes than the winner.
The audit stops (short of a full hand count) only when there is sufficiently strong evidence that every winner beat every loser. 

\emph{Comparison audits} involve manually checking the voting system's interpretation of the votes on physically identifiable subsets of ballots.
They require the voting system to export vote tallies for physically identifiable subsets of ballots, so that the votes on those ballots can be tallied by hand and compared to the voting system's tallies.
They also require checking that the reported subtotals yield the reported contest results, and that the subtotals account for all ballots cast in the contest.
They generally also require ballot manifests.

A comparison audit that checks the voting system's interpretation of individual ballots is a \emph{ballot-level comparison} RLA.
Ballot-level comparison RLAs are more efficient than batch-level comparison RLAs and ballot-polling RLAs in that they generally require examining fewer ballots when the reported outcome is correct.
However, they have higher set-up costs and require more data export from the voting system: they need a \emph{cast vote record} or \emph{CVR} for
each physical ballot, a way to locate the CVR for each physical ballot, and \emph{vice versa}.
(A CVR is the voting system's interpretation of voter intent for a given ballot.)
Relying on more general results in \cite{Stark2009b} for batch-level comparison audits, \cite{Stark2010} 
developed a sequential ballot-level 
comparison RLA method that results in particularly simple calculations.

\emph{Transitive audits}  (\cite{calandrino2007machine,LindemanStark2012}) involve auditing an unofficial system that is easier to audit than the official system.
If the two systems agree who won, an audit that provides strong evidence that the unofficial system found the correct winner(s) transitively provides strong evidence that the official system did also; and if the audit of the unofficial system leads to a full manual tally, the outcome of that tally can be used to correct the official result.
A transitive audit does not confirm that the official system tallied votes correctly; indeed, the two systems might disagree about the interpretation of every ballot, but still agree who won.

Indian EVMs do not create CVRs, but---if the VVPATs are organized appropriately---they can still be audited using a transitive ballot-level comparison audit.
CVRs can be constructed for EVMs by sorting the VVPATs into bundles that 
(purportedly) show the same voter preferences, counting the number of VVPATs in each batch, and labeling each bundle with the number of ballots and the voter preferences it purports to contain.
A report of the bundle labels, the number of VVPATs in each bundle, and the reported voter preference for the bundle amounts to a CVR for every VVPAT.
Such a report in effect combines a \emph{ballot manifest} (\cite{LindemanStark2012}) and a commitment to a cast vote record for every ballot, implied by the label of the bundle the ballot is in. 
We shall call such a report a \emph{preference manifest}.

If ballots are sufficiently simple---\emph{e.g.}, if each contains only one contest, as in India---sorting ballots by voter preference can be practical.
Indeed, this is how ballots are tallied manually in Denmark: on election night, ballots are sorted within polling places according to the voter's party preference.
The following day, ballots are sorted further according to the voter's candidate preference, to produce homogeneous bundles of ballots, each labeled with the number of ballots and the voters' preference.

Such sorting-based CVRs were the basis of a RLA in Denmark (\cite{Schurmann2016}). 
The sorting might be manual, as it is in Denmark, but it could be automated partly or entirely.
(Sorting may also increase vote anonymity by breaking any link between voter and ballot.)
When the official tallying process itself is based on creating and counting the homogeneous bundles, as it is in Denmark, the audit is a direct audit of the voting system.
If the sorting is conducted independently of the tabulation, as it would be if India were to sort the paper ballots to produce a preference manifest, the resulting audit is a transitive audit.

The first step of a ballot-level comparison RLA is to verify that the CVRs are consistent with the reported results: that applying the social choice function to the vote subtotals implied by the sizes of the bundles and the votes they purport to contain produces the same set of winners. If the preference manifest does not produce the same set of winners as reported, the audit should not continue: there is a serious problem. 
The audit should also check that the number of CVRs for each contest does not exceed the number of ballots cast in the contest, which should be determined without reliance on the voting system (\cite{banuelos2012limiting}).
If the preference manifest passes these checks, the audit can begin to select ballots at random to check the accuracy of the CVRs implied by the preference manifest against a manual reading of voter intent from each paper ballot.

\cite{Kroll2014} present a method for reducing the workload in auditing multi-level elections, inspired by the US Electoral College. 
They show that to achieve an overall confidence that a party or coalition secured the majority of seats, the individual constituencies can sometimes be audited to lower confidence levels.
They provide a constraint optimization program describing the set of feasible solutions ({\it i.e.} those that constitute a sufficient audit) and 
a number of methods for finding the optimal solution.
In India's electoral system, as in many other parliamentary democracies, every constituency has equal weight.  

\section{Auditing individual constituencies using extant methods}\label{secApplying}

This section discusses how existing methods for RLAs apply to Indian elections.  
We consider auditing individual constituencies rather than the entire election; Section~\ref{secOverall} shows how to combine audits of constituencies to audit an entire contest. 

India's voting system currently does not support ballot-level comparison audits, but, as described above, if procedures were added to sort the paper ballots and to report a preference manifest, transitive ballot-level comparison audits would be possible.
Because ballots in India are simple---a single selection in a single contest---such sorting is feasible.

Ballot-polling audits could be used in India without requiring sorting the ballots or modifying the voting system, if ballot manifests were available (see section~\ref{subsecBPMul}). 
The calculations for BRAVO (\cite{Lindeman2012}) and the ballot-polling method in~\cite{LindemanStark2012} are simple enough to do with a pencil and paper or hand calculator, and have open-source online tools by Stark: \url{https://www.stat.berkeley.edu/~stark/Vote/ballotPollTools.htm}

When the election outcome is correct, ballot-level comparison audits generally require inspecting fewer ballots than ballot-polling audits. 
(Because they are RLAs, when the outcome is incorrect, both methods have a large chance of requiring a full manual tally.)
The advantage grows as the margin shrinks: as a rule of thumb, workload increases inversely with the reported margin for ballot-level comparison audits, and increases inversely with the square of the actual margin for ballot-polling audits. 
However, preparing for a ballot-level comparison audit is harder, because it requires CVRs linked to the corresponding physical ballots. While they may require inspecting more ballots, the simplicity of ballot-polling audits may offset the work of examining more paper, unless the margin is very small.

\subsection{Transitive Ballot-Level Comparison RLA}\label{subsecSorted}

Ballot-level comparison RLAs were introduced by~\cite{Stark2010} who provides online tools at 
\url{https://www.stat.berkeley.edu/~stark/Vote/auditTools.htm}; see also \cite{LindemanStark2012}.
Ballot-level comparison audits require a way to find the CVR corresponding to each paper ballot, and vice versa. 
The EVMs currently used in India do not provide CVRs at all. 

However, sorting ballots into groups according to the vote (if any) that they are reported to show in effect provides a CVR for each ballot through a preference manifest that lists the bundles of ballots, the number of ballots in each bundle, and the (single) preference that every ballot in the bundle is supposed to show.
This approach was used by \cite{Schurmann2016} to audit an election in Denmark.

In Denmark, ballots are manually sorted into bundles with homogeneous voter intent, but sorting could be automated with relatively simple equipment, possibly something similar to the system used in South Korea.
Whether it is worth the effort to sort the ballots depends on the margin in the contest: if the margin is wide, it will be less expensive to use ballot polling, but if the margin is very narrow, the cost of sorting---whether manual or automated---may reduce the sample size required to confirm the outcome by orders of magnitude. 

 \subsubsection{Classifying CVR errors}
 \cite{Stark2009b} reviews a number of methods to test the hypothesis that any loser received more votes than any winner by comparing hand counts of votes in randomly selected batches of ballots to the machine counts of the votes on the same ballots.
 The methods apply to arbitrarily small batches, including batches consisting of a single ballot; that is, to ballot-level comparison audits.
 \cite{Stark2010} elaborated on one of those methods, which relies on the Kaplan-Markov inequality.
 By introducing a taxonomy of discrepancies, the arithmetic can be simplified to the point that a pencil and paper suffice, while rigorously maintaining the risk limit.
 That ``super-simple simultaneous single-ballot'' method was further simplified by \cite{LindemanStark2012}, and is the basis of pilot audits in Denmark, California, Colorado, Indiana, Michigan, New Jersey, and Virginia, and of the statutory risk-limiting audits in most Colorado counties.
 
 \cite{stark2014verifiable} presented a ballot-level comparison RLA method based on the Kaplan-Wald inequality, which has some advantages over the Kaplan-Markov inequality.
 In this paper, we use the Kaplan-Markov method as simplified by \cite{LindemanStark2012}, because it has been used more widely.
 We shall refer to it as the \emph{LSKM method}.
 It is straightforward to modify the procedures below to use the 
 method of \cite{stark2014verifiable} instead.
 
 The LSKM method is \emph{sequential}: it involves examining more and more ballots selected at random until either there is strong evidence that the reported winners really won, or until there has been a full hand count and the correct outcome is known. 
 Conceptually, after examining one or more ballots, one calculates a sequentially valid\footnote{%
 \emph{Sequentially valid} means that the chance that the infimum of the $P$-value over all sample sizes is less than or equal to $p$ is less than or equal to $p$ if the null hypothesis is true.
 In contrast, standard hypothesis tests are designed for sample sizes that are fixed ahead of time: expanding the sample and re-calculating the $P$-value for such tests generally produces type~I error rates far larger than the nominal significance level, because it does not account for multiplicity.
 }
 $P$-value of the hypothesis that the outcome is wrong.
 If that $P$-value is less than or equal to the risk limit, the audit stops; otherwise, more ballots are audited
 and the sequential $P$-value is updated.
 The method presented below to check the overall electoral outcome involves combining the $P$-values for individual constituencies.
 
 If the audit does lead to a full hand tally in a constituency, the reported results are replaced by the results according to that full hand tabulation.
 Election officials may elect to terminate the audit and conduct a full hand count at any time, for instance, if they estimate that the cost of additional sampling will exceed the cost of a full manual tally.
 
 The LSKM method involves classifying discrepancies between the CVR and a manual reading of voter intent from the paper ballot:
 \begin{itemize}
     \item If correcting the CVR would reduce the margin between any (reported) winner and any (reported) loser by two votes, the discrepancy is a \emph{2-vote overstatement} (the number of 2-vote overstatements is denoted $o_2$).
     \item If not, but if correcting the CVR would reduce the margin between any winner and any loser by one vote, the discrepancy is a \emph{1-vote overstatement} (the number of 1-vote overstatements is denoted $o_1$). 
     \item If not, but if correcting the CVR would not increase the margin between every winner and every loser, the discrepancy is a \emph{neutral error}.
     (Neutral errors do not enter the stopping rule explicitly.)
     \item If not, but if correcting the CVR would increase the margin between every winner and every loser by at least one vote, and increase the margin between some winner and some loser by exactly one vote, the discrepancy is a \emph{1-vote understatement} (the number of 1-vote understatements is denoted $u_1$).
     \item If correcting the CVR would increase the margin between every winner and every loser by two votes, it is a \emph{2-vote understatement} (the number of 2-vote understatements is denoted $u_2$).
 \end{itemize}
 
 Two-vote overstatements should be rare if the voting system is working correctly: they involve mistaking a vote for a loser as a vote for a winner.
 Two-vote understatements should be even rarer---and are typically mathematically impossible. 
 For instance, in a plurality, vote-for-one contest with three or more candidates, two-vote understatements are impossible, because they would require having mistaken a valid vote for the winner as a valid vote for every losing candidate.
 
 We assume that there is a trustworthy upper bound on the total number of ballots cast, for instance, from pollbooks or from information about the number of eligible voters. 
 A preliminary check should ensure that the preference manifest does not list more ballots than that upper bound: if there are more ballots listed than can exist, there is a serious problem that the audit cannot address by itself.\footnotemark
 
 \footnotetext{Prof Sandeep Shukla of IIT Kanpur has pointed out that the current Indian VVPAT design does not protect against the EVM adding electronic votes and corresponding VVPATs when the voter is not looking, because there is no publicly observable mechanism to ensure that at most one VVPAT is inserted into the box per voter.  This needs to be addressed by improving the physical design in a way that is out of the scope of this paper.}
 
 In the sorted-ballot method described above,
  \begin{itemize}
     \item a \emph{2-vote overstatement} occurs if we find a vote for a reported loser in the reported winner's pile;
     \item a \emph{1-vote overstatement} occurs if we find a vote for a different reported loser in a reported loser's pile;
     \item \emph{neutral errors} don't occur;\footnotemark 
     \item a \emph{1-vote understatement} occurs when there are at least three candidates and we find a vote for the reported winner in a reported loser's pile;

     \item a \emph{2-vote understatement} occurs only when there are exactly two candidates, and we find a vote for the reported winner in the reported loser's pile;
     \item if a pile turns out to be \emph{smaller} than reported, the discrepancy can be addressed using the ``phantom to zombie'' approach of~\cite{banuelos2012limiting}.
     \item if a pile turns out to be \emph{larger} than reported, then some other pile must be \emph{smaller} than reported, and the ``phantom to zombie'' approach of~\cite{banuelos2012limiting} will still ensure that the risk is controlled conservatively.
 \end{itemize}

 \footnotetext{Indian EVMs (as far as we know) do not produce blank votes.  However, if they did they could be accommodated easily.  A 1-vote overstatement occurs if we find a blank vote in the reported winner's pile.  A netural error would occur when there were at least three candidates and we found a blank vote in a reported loser's pile.       A one-vote understatement would occur when there were exactly two candidates and we found a blank vote in the reported loser's pile.}
 
 There are sharper ways to treat discrepancies than to use these categories (in particular, keeping track of \emph{which} margins are affected by each discrepancy can reduce the number of ballots the audit inspects; see \cite{Stark2010}).
 However, the bookkeeping is more complex. 
Categorizing discrepancies this way makes the calculations simple enough to do with a pencil and paper (aside from calculating 5 constants involving logarithms, which can be done once and for all and verified by anyone).
 
 \subsubsection{Calculations}
 Let $n$ denote the current sample size and $\alpha$ the risk limit. Fix $\gamma \ge 1$.
 The LSKM method stops auditing (and concludes that the reported winners really won) if
 \begin{equation} \label{eq:LSKM}
 n \geqslant \frac{2\gamma}{\mu} \left ( o_1\log \left (\frac{1}{1-\frac{1}{2\gamma}}\right )+ o_2\log \left (\frac{1}{1-\frac{1}{\gamma}} \right )- u_1\log \left (1+\frac{1}{2\gamma} \right ) - u_2\log \left (1+\frac{1}{\gamma} \right ) - \log(\alpha) \right ).
 \end{equation}
In this expression, $\mu$ is the \emph{diluted margin}, the smallest difference in votes between any winner and any loser, divided by the total number of ballots in the population from which the sample is to be drawn, including ballots with invalid votes. 
The constant $\gamma \ge 1$ is the \emph{error inflation factor}, which controls the operating characteristics of the LSKM method when errors are observed: the larger $\gamma$, the fewer additional ballots need to be audited if a 2-vote overstatement is observed, but the smaller $\gamma$ is, the fewer ballots need to be audited if no 2-vote overstatements are observed. Because two-vote overstatements should be rare, taking $\gamma$ slightly larger than 1 should suffice. For $\gamma$ exactly equal to 1, then if the audit finds even one 2-vote overstatement, the audit will not terminate without a full hand count. \cite{LindemanStark2012} suggest using $\gamma = 1.03905$, which makes the ``cost'' of a 2-vote overstatement 5 times larger than the ``cost'' of a 1-vote overstatement, where ``cost'' means the number of additional ballots that must be audited to attain the risk limit.
Any value of $\gamma \ge 1$ gives a risk-limiting audit, but $\gamma$ must be chosen before any ballots have been audited.

\subsection{Ballot-polling RLA using BRAVO}\label{subsecBPMul}
The BRAVO ballot-polling RLA by \cite{Lindeman2012}
can be applied immediately to constituencies in India.
In the Indian scenario, we have only one winner per constituency and one candidate per ballot.  
For each loser $\ell$, the null hypothesis $H_{0w\ell}$ states that $w$ did not get more votes than $\ell$, that is, that the true outcome was a tie or that $\ell$ actually won. 
BRAVO uses Wald's Sequential Probability Ratio Test 
\cite{Wald1945}
to test all the null hypotheses simultaneously.\footnotemark

The audit begins by choosing the risk limit $\alpha$, the maximum probability of stopping short of a full manual tally if a full manual tally would show a different electoral outcome.
It also requires the reported vote totals for each candidate,\footnote{%
There are other ballot-polling methods that do not use the reported results at all.
} but no other data from the voting system.

For every apparent loser $\ell$, define the \emph{conditional vote share $s_{w\ell}$ }:

\begin{equation}
   s_{w\ell} \equiv \frac{v_w}{v_w+v_\ell}
\end{equation}

Here, $v_w$ and $v_\ell$ are the reported vote totals for the winner $w$ and the loser $\ell$ respectively.
If the reported vote tally is correct,
the chance that a randomly selected ballot shows a vote for $w$, given that it shows a vote for either $w$ or $\ell$, is $s_{w\ell}$.

BRAVO maintains a test statistic $T_{w\ell}$ for each reported (winner, loser) pair.
In Indian elections, there is only one reported winner $w$ per constituency, so
this amounts to a test statistic for each reported loser $\ell$.
Null hypothesis $H_{0w\ell}$ is rejected if:
\begin{equation} \label{eq:19}
T_{w\ell} \geqslant \frac{1}{\alpha}.
\end{equation}
If the null hypotheses for all apparent losers $\ell \in L$ are rejected, we stop the audit and declare the announced outcome correct.

At any time, for example if the audit is expected to take more time than simply counting the ballots, we can stop the audit and perform a full manual recount. The algorithm runs as follows:

\begin{algorithm}[H]
\caption{BRAVO with protection against manifest errors.\\
This is a simplified version of BRAVO that assumes the contest has only one winner, and that there can be at most at most one valid vote per ballot. 
It incorporates the ``phantom to zombie'' method of \cite{banuelos2012limiting} for dealing with errors in the ballot manifest.}\label{bravo}
\begin{algorithmic}[1]
\INPUT Risk Limit $\alpha$; ballot manifest, announced winner $w$, losers $L$ and corresponding weighted vote shares $s_{w\ell}$ for each $\ell \in L$.
Upper bound $N$ on the number of ballots, where $N$ is at least as large as the number of ballots listed in the manifest.
Work threshold $K \le N$.
\State $\text{Simultaneous probability ratio: $\forall \ell \in L$}: T_{w\ell} \gets 1$
\State $\text{Number of audited ballots: } n  \gets 0$
\While {not every $H_{0w\ell}$ has been rejected}
\State $\text{Generate a random number \textit{i} between 1 and $N$}$
\State $\text{Look up the $i$th ballot in the ballot manifest and (attempt to) retrieve it}$
\If {$\text{Ballot \textit{i} is not in the ballot manifest or cannot be found}$}
\State $\text{For every $\ell \in L$, multiply } T_{w\ell} \text{ by } 2(1-s_{w\ell})$
\ElsIf {$\text{Ballot \textit{i} shows a vote for the winner $w$}$}
\State $\text{For each $\ell \in L$, multiply } T_{w\ell} \text{ by } 2s_{w\ell}$
\ElsIf {$\text{Ballot \textit{i} shows a vote for loser $\ell \in L$}$}
\State $\text{Multiply } T_{w\ell} \text{ by } 2(1-s_{w\ell})$
\EndIf
\If {$T_{w\ell} \geqslant \frac{1}{\alpha}$}
\State \text{Reject } $H_{0w\ell}. \text{ Withdraw the pair $(w,l)$ from the audit.}$
\EndIf
\If {\text{the number of ballots inspected exceeds $K$, or optionally at any time}} 
\State {\bf STOP} the audit and perform a full manual recount.
\EndIf
\EndWhile
\State Declare election outcome correct---all null hypotheses have been rejected.
\end{algorithmic}
\end{algorithm}

At any stage, $P = \max_{\ell \in L} 1/T_{w\ell}$ is a conservative sequential $P$-value for the hypothesis that the reported winner $w$ did not actually win
the constituency.

\subsubsection{Number of votes to be audited}

Consider an example of a 3-candidate contest with a single plurality winner. The candidates are Ramith, Shyam and Priya. Their respective shares is recorded in the following table:


\begin{center}
\begin{tabular}{|l|l|l|}
\hline
\textbf{Ramith} & \textbf{Shyam} & \textbf{Priya} \\
\hline
$20,000$ & $30,000$ & $50,000$ \\
\hline
\end{tabular}
\end{center}

In this case, the winner is Priya. Let us denote the winner-loser pairs as $(p,r)$ for Priya and Ramith and $(p,s)$ for Priya and Shyam. The weighted vote shares are:

\begin{equation*}
s_{pr} = \frac{v_p}{v_p+v_r} = \frac{50}{50+20} = 0.714
\end{equation*}

\begin{equation*}
s_{ps} = \frac{v_p}{v_p+v_s} = \frac{50}{50+30} = 0.625
\end{equation*}

We set the risk limit $\alpha$ at $5\%$.
Every time the audit selects a ballot that shows a vote for Priya we multiply $T_{pr}$ by $\frac{0.714}{0.5}=1.428$ and $T_{ps}$ by $\frac{0.625}{0.5}=1.25$. 
Therefore, the minimum sample size $n$ to attain a risk limit $\alpha = 0.05$
satisfies

\begin{equation*}
1.428^{n} \geq 20
\end{equation*}
and
\begin{equation*}
1.25^{n} \geq 20
\end{equation*}

The smallest such $n$ is $n=14$. 
Hence, we need to audit at least 14 ballots---if they all show up votes for Priya, BRAVO will confirm the election outcome at risk limit 5\%.

If the reported election results were accurate, on average we would see $50\%$ of ballots for Priya, $30\%$ for Shyam and $20\%$ for Ramith.
\cite{Lindeman2012} describe how to find the \emph{Average sample number (ASN)},
the expected sample size necessary to reject all the null hypotheses, assuming the reported results are indeed correct.  
Stark's online ballot-polling tool shows an ASN of 123 for this example.

\subsection{Improved methods for single-constituency RLA's}
There have been numerous improvements to the efficiency of Risk-Limiting Audits, any of which could easily apply to India's simple electoral system.  See for example \url{https://github.com/pbstark/S157F17/blob/master/kaplanWald.ipynb} and \url{https://github.com/pbstark/S157F17/blob/master/pSPRTnoReplacement.ipynb}.

The next section explains how to audit the overall parliamentary winner by an efficient combination of single-constituency audits. 
It requires independent, sequentially valid $P$-values $\{P_i\}$ for the hypotheses that the reported outcome in constituency $i$ is incorrect.
It does not require the $P$-values to be obtained using the same method.
For instance, some constituencies could use ballot polling and others could use
transitive ballot-level comparison audits.

\section{Auditing the overall parliamentary winner}\label{secOverall}
A party or a coalition needs a majority of the seats in the Lower House of Parliament to form a new government. 
The total number of seats is 543, so to win, a party or coalition needs at least 272 seats. 
The audit needs to confirm that the reported winning party or coalition truly won at least 272 seats.
(The particular seats the reported winner won is immaterial to whether they won overall.)
If party $w$ supposedly won $M \ge 272$ constituencies, then for a different 
party to have won in fact, the reported outcome must be wrong in at least $m = M-271$ of the constituencies that $w$ supposedly won.
Note that this condition is necessary but not sufficient for the parliamentary outcome to be wrong: if $w$ in fact won some constituencies it was reported to have lost, the outcome could be wrong in $m$ constituencies $w$ supposedly won and yet $w$ could still be the overall winner.
However, if the audit provides strong evidence that there is no set of
$m$ constituencies $w$ reportedly won for which $w$ did not actually win,
$w$ must be the overall winner.

Let $W$ denote the set of constituencies $w$ reportedly won. 
Then $|W| \ge 272$ and $m = |W|-271$, where $|W|$ denotes the cardinality of
the set $W$.
If there is no set of constituencies $C \subset W$ with $|C|=m$ for which
$w$ lost in \emph{every} $c \subset C$, $w$ really won overall.

Let $\alpha$ denote the overall risk limit, and let $P_i$ denote a
$P$-value for the hypothesis that the reported outcome in constituency $i$
is wrong.
We suppose that the audits in different constituencies rely on independently selected random samples of ballots, so
the $P$-values $\{P_c\}$ are independent random variables.
If the reported outcome in constituency $c$ is incorrect, the probability distribution of $P_c$ is stochastically dominated by a uniform distribution.
That is, $\Pr \{P_c \le p \} \le p$ if the reported outcome in constituency $c$ is
wrong. 

\emph{Fisher's combining function} for a set of $P$-values $\{P_c \}_{c\in C}$
is
\begin{equation}
    X^2(C) \equiv -2 \sum_{c \in C} \ln P_c.
\end{equation}
If the $P$-values $\{P_c\}$ are independent and all the null hypotheses are true, the probability distribution of $X^2(C)$ is stochastically smaller than a
chi-square distribution with $2|C|$ degrees of freedom.\footnote{%
See, e.g., \cite{ottoboni2018risk}.
}
That is, if the reported outcome in every constituency $c \in C$ is wrong,
\begin{equation}
    \Pr \{ X_{2|C|}^2 \ge \chi_{2|C|}^2(1-\alpha) \} \le \alpha,
\end{equation}
where $\chi_{2|C|}^2(1-\alpha)$ is the $1-\alpha$ quantile of the chi-square
distribution with $2|C|$ degrees of freedom.

The overall strategy for auditing the parliamentary outcome is thus as follows:

\begin{enumerate}
    \item Select an overall risk limit $\alpha$ for the parliamentary outcome.
    \item Definitions:
        \begin{itemize}
            \item $W$ denotes the constituencies the reported winning party allegedly won
            \item $m \equiv |W|-271$
            \item $\mathcal{W}_m$ denotes the set of all subsets of $W$ with cardinality $m$
            \item $\chi^2 \equiv \chi_{2m}^2(1-\alpha)$ denotes the $1-\alpha$ quantile of the chi-square distribution with $2m$ degrees of freedom
            \item for any set $C$ of constituencies, $X^2(C) \equiv -2 \sum_{c \in C} \ln P_c$, where $P_c$ is the $P$-value of the hypothesis that the reported winner in constituency $c$ did not really win, based on the audit sample selected from constituency $c$ so far. Before any data are collected from $c$, $P_c = 1$.
            \item For any collection $\mathcal{U}$ of sets of constituencies, define $\mathcal{U}(c) \equiv \{U \in \mathcal{U}: U \ni c \}$, all sets of constituencies in $\mathcal{U}$ that contain $c$.
        \end{itemize}
    \item Initialization: Set $\mathcal{U} = \mathcal{W}_m$. Select an initial sample size $n_c$ for each constituency $c$, and draw the initial sample. It is permissible, but not advisable, to let $n_c = 0$ in any constituency $c \notin W$.
    \item Audit: While $\mathcal{U}$ is not empty:
        \begin{enumerate}
            \item If there is any $C \in \mathcal{U}$ for which every $c \in C$ has been fully hand counted and the hand count has shown that the reported winner was incorrect in every $c \in C$, stop and perform a full hand count of the entire election.
            \item For every constituency $c$ that has been fully hand counted, if the hand count confirms the reported outcome in $c$, $\mathcal{U} \gets \mathcal{U} \setminus \mathcal{U}(c)$
            \item $\mathcal{U} \gets \mathcal{U} \setminus \{ C \in \mathcal{U}: X^2(C) \ge \chi^2 \}$
            \item Increase $n_c$ in one or more constituencies $c \in \cup_{U \in \mathcal{U}} \cup_{u \in U} u$ and inspect the additional ballots.\footnote{%
            The rule for increasing sample sizes could be as simple as ``increase every $n_c$ by 25\%,'' or it could be designed to minimize the expected total amount of auditing required, for instance, by preferentially increasing the sample size in constituencies with large margins and taking into account differences in auditing methods in different jurisdictions (ballot polling versus transitive ballot-level comparison). 
            All else equal, when the outcome is correct, auditing an additional ballot is expected to decrease the $P$-value more the larger the true margin is. Similarly, all else equal, auditing an additional ballot in a jurisdiction conducting a transitive ballot-level comparison RLA is expected to decrease the $P$-value more than auditing an additional ballot in a jurisdiction conducting a ballot-polling RLA.}
        \end{enumerate}
    \item If the loop terminates with $\mathcal{U} = \emptyset$, the audit has confirmed the parliamentary outcome at risk limit $\alpha$.
\end{enumerate}

\noindent
\textbf{Proof that the algorithm above 
is an RLA for the parliamentary outcome.}
We need to show that if the parliamentary outcome is wrong, the chance that the audit stops without a full manual tally of every constituency is at most $\alpha$.
If the parliamentary outcome is wrong, the reported winner is wrong in every $c \in C$ for some $C \in \mathcal{W}_m$.
Suppose there is such a $C$.
If the audit leads to hand counting every $c \in C$, step~(4a) ensures that there will be a full hand count of the entire election.
Therefore, there will be a full hand count unless $C$ is removed from $\mathcal{U}$.
There are two places that sets of constituencies can be removed from $\mathcal{U}$: step~(4b) and step~(4c).
Step~(4b) cannot remove $C$ from $\mathcal{U}$, because, by assumption, handcounting any $c \in C$ would not confirm the reported outcome in $c$.
Therefore, the chance that $C$ is \emph{not} fully hand counted is at most the chance that step~(4c) removes $C$ from $\mathcal{U}$. But, by construction (through Fisher's combining function applied to the independent constituency-level $P$-values), that chance is not larger than $\alpha$.
If there is more than one $C \in \mathcal{W}_m$ for which every reported outcome is wrong, the audit must erroneously remove \emph{all} of them at step~(b). But the chance of erroneously removing all of them cannot be larger than chance of removing any one of them individualy, which is in turn at most $\alpha$.

\section{Conclusion and future work}

We have presented an approach to conduct
risk-limiting audits of the national outcome of Indian elections
by combining audits conducted in different constituencies using independent samples. 
Within a given constituency, the audit could use ballot polling, or---with
an initial step of sorting VVPATs---transitive ballot-level comparisons.
The $P$-values in different constituencies are combined using Fisher's combining function, for a collection of sets of constituencies.
The collection is constructed in such a way that for the reported parliamentary outcome to be wrong, the reported outcome must 
be wrong in \emph{every} constituency in at least one of the sets.
If there is strong statistical evidence that there is no set of constituencies 
in the collection for which every reported outcome is wrong, that confirms the national parliamentary outcome.
The multi-level structure of Indian parliamentary elections makes it possible to have high confidence in the overall parliamentary outcome without necessarily auditing every constituency to a low risk limit.  
In future research we will address how to schedule increases in sample sizes in different constituencies to minimize the total expected workload, taking into account the reported margins in different constituencies and the auditing methods used in different constituencies.

\section*{Acknowledgments}
Many thanks to Archanaa Krishnan, Chittaranjan Mandal, Sandeep Shukla, Peter Stuckey and Poorvi Vora for valuable suggestions on this work.

\printbibliography

\end{document}